\documentclass[ aps, showpacs, showkeys, nofootinbib, floatfix, superscriptaddress]{revtex4}

\usepackage{amsfonts}

\usepackage{amssymb}
\usepackage{amsmath}
\usepackage{graphicx}
 \usepackage{amsmath}
\begin{document}

\title{  Cosmic constraint on massive neutrinos in viable f(R) gravity with producing $\Lambda$CDM background expansion}

 \author{Jianbo Lu}
 \email{lvjianbo819@163.com}
 \affiliation{Department of Physics, Liaoning Normal University, Dalian 116029, P. R. China}
\author{Molin Liu}
 \affiliation{Department of Physics, Xinyang Normal University, Xinyang 116029, P. R. China}
\author{Yabo Wu}
 \affiliation{Department of Physics, Liaoning Normal University, Dalian 116029, P. R. China}
\author{Yan Wang}
 \affiliation{Department of Physics, Liaoning Normal University, Dalian 116029, P. R. China}
\author{Weiqiang Yang}
 \affiliation{Department of Physics, Liaoning Normal University, Dalian 116029, P. R. China}

\begin{abstract}
  Tensions between several cosmic observations  were found  recently, such as the inconsistent values of $H_{0}$ (or $\sigma_{8}$) were indicated by the different cosmic observations. Introducing the massive neutrinos in $\Lambda$CDM could potentially solve the tensions. Viable $f(R)$ gravity  producing $\Lambda$CDM background expansion with massive neutrinos is investigated in this paper. We fit the current observational data: Planck-2015 CMB, RSD, BAO  and SNIa to constrain the mass of  neutrinos in viable $f(R)$ theory. The constraint results  at 95\% confidence level are: $\Sigma m_\nu<0.202$ eV for the active neutrino case, $m_{\nu, sterile}^{eff}<0.757$ eV with $N_{eff}<3.22$ for the sterile neutrino case. For the effects by the mass of neutrinos, the constraint results  on model parameter at 95\% confidence level  become $f_{R0}\times 10^{-6}> -1.89$  and $f_{R0}\times 10^{-6}> -2.02$ for two cases, respectively. It is also shown that the fitting values of several parameters much depend on the  neutrino properties, such as the cold dark matter density, the cosmological quantities at matter-radiation equality, the neutrino density and the fraction of baryonic mass in helium. At last, the constraint result shows that  the tension between direct and CMB measurements of $H_0$ gets slightly weaker in the viable $f(R)$ model than that in the base $\Lambda$CDM model.

\end{abstract}



\maketitle

\section{$\text{Introduction}$}

 { The base 6-parameter $\Lambda$CDM ($\Lambda$-Cold-Dark-Matter) model  is the most popular one to interpret the accelerating expansion of universe. This model is favored by most "observational probes", though it exists the fine-tune problem and the coincidence problem in theory. However,  some tensions were found recently between the cosmic observations when one fitted observational data to this model.  For example, the tension is found for estimating the values of  $H_{0}$: a lower value of $H_{0}=67.3\pm1.0$ is provided  by Planck-CMB experiment with an indirect estimate on $H_{0}$ \cite{Planck2015-1}, but a higher value of $H_{0}=74.3\pm2.1$ is obtained by SST direct measurements of $H_{0}$ \cite{H0-high1}; this tension also exists  between the Planck-CMB experiment and the rich cluster counts, as they provide the the inconsistent value of $\sigma_{8}$  \cite{Planck2015-1,tension-low-sigma8}.

 The studies on these tensions are important, since any evidence of a tension may be useful to  search new physics. One possible interpretation to above tension is that the base 6-parameter $\Lambda$CDM model is  incorrect or should be extended.   Ref.  \cite{Planck2015-1} shows that, introducing $\sum m_{\nu}$ or introducing $N_{eff}$ solely in $\Lambda$CDM model can not resolve the above tensions, but the tensions could be solved in the $\Lambda$CDM with including  both $\sum m_{\nu}$ and  $N_{eff}$ or with including  the massive sterile neutrinos  $ m_{\nu, eff}^{sterile}$.  Here $\sum m_{\nu}$  denotes  the total mass of  three species of degenerate massive active neutrinos, and  $N_{eff}$  denotes the effective number of relativistic degrees of freedom, which relates to the neutrinos and the extra massless species.   Combined analysis of cosmic data in other references also indicate the existence of the massive neutrinos, for examples, joint analysis from CMB and BAO (baryon acoustic oscillation)  \cite{massive-neutrinos1,massive-neutrinos2}, from solar and atmospheric experiments \cite{massive-neutrinos3,massive-neutrinos4,massive-neutrinos5}, or from the reactor neutrino oscillation anomalies \cite{massive-neutrinos6,massive-neutrinos7}, etc..

  Investigating other scenarios to solve the above tensions and restricting the mass of neutrinos in different scenarios are significative.  Ref. \cite{favor-fr-neutrinos} shows that possible discovery of sterile neutrino with mass $ m_{\nu,sterile}^{eff}\approx 1.5 eV$, motivated by various anomalies in neutrino oscillation experiments, would favor cosmology based on $f(R)$ gravity rather than the standard $\Lambda$CDM. In addition, one knows that plenties of   functions  $f(R)$ of Ricci scalar $R$ \cite{fr-review,fr-1,xu-fr3-GW,fr-4,fr-5,fr-6,fr-7,fr-8,fr-9,fr-10,fr-11,fr-12,fr-13,fr-14,fr-15,fr-16,fr-17} are presented to modify the Einstein's gravity theory, in order to solve  puzzles in general relativity. But several forms of  $f(R)$ are then found to be nonphysical, since they can not describe the expansion of universe in matter-dominated time \cite{fr-nonphysical,fr-nonphysical1}. So, studies on observationally viable $f(R)$ theories are necessary. One of the viable $f(R)$ theories has been studied in Refs. \cite{viable-fr-lcdm,viable-fr-lcdm1},  where the $f(R)$ theory can realize the most popular $\Lambda$CDM universe at background-dynamics level, while the effects of large scale structure with the cosmological perturbation theory in this $f(R)$ model are different from that in the $\Lambda$CDM.  In this paper, we investigate the behaviors of massive neutrinos in observationally viable $f(R)$ theories with producing the $\Lambda$CDM background expansion history.

\section{$\text{Viable f(R) gravity theory producing  $\Lambda$CDM-background expansion}$}

The action of $f(R)$ modified gravity theory is written as
\begin{equation}
I=\int d^{4}x\sqrt{-g}[\frac{1}{16\pi G}f(R)+\mathcal{L}_{u}].
\end{equation}
$\mathcal{L}_{u}$ is the Lagrangian density of universal matter including the radiation and the pressureless matter (baryon matter plus cold dark matter). Using the variation principle, one gets
\begin{equation}
f_{R}R_{\mu\nu}-\frac{1}{2}f(R)g_{\mu\nu}+(g_{\mu\nu}\Box -\nabla_{\mu}\nabla_{\nu})f_{R}=8\pi G T_{\mu\nu}.
\end{equation}
 $f_{R}=\frac{d f(R)}{d R}$, $R_{\mu\nu}$ and $T_{\mu\nu}$   denote the Ricci tensor
 and  the energy-momentum tensor  of universal   matter, respectively. For a universe  described by  metric $g_{\mu\nu}=diag (-1, a(t)^{2}, a(t)^{2}, a(t)^{2})$,  the dynamical evolutionary equations of universe in $f(R)$ theory are
\begin{equation}
3f_{R}H^{2}=\frac{f_{R}R-f}{2}-3Hf_{RR}\dot{R}+8\pi G(\rho_{m}+\rho_{r})
\end{equation}
\begin{equation}
2f_{R}\dot{H}=H\dot{f}_{R}-\ddot{f}_{R}-k^{2}[\rho_{m}+\frac{4\rho_{r}}{3}].
\end{equation}

As shown in Refs. \cite{viable-fr-lcdm1},  the viable $f(R)$ theory which realize the  popular $\Lambda$CDM universe at background-dynamics level does not have an analytical expression of $f(R)$ to describe a physical universe from the radiation-dominate epoch to the late-time acceleration, but it really has the analytical solutions of $f(R)$ in different evolutional epochs of the universe. Concretely, Ref. \cite{viable-fr-lcdm1} gives the forms of $f(R)$ in two cases: one describes the evolution of the $\Lambda$CDM background from the radiation-dominate epoch to the matter-dominate epoch, and the other one represents the evolution of the $\Lambda$CDM background from the matter-dominate era to the future expansion. In this paper we focus on studying the $f(R)$ function with producing  $\Lambda$CDM background expansion from the  matter-dominate epoch to the late-time acceleration\footnote{An accelerating cosmological model can be used to interpret the current observations. And for the $\Lambda$CDM background expansion from the  matter-dominate epoch to the late-time acceleration, $R$ can be written by $R=3\Omega_{m}a^{-3}+12\Omega_{\Lambda}=3\Omega_{m}a^{-3}+4\Lambda$.}, which has the form as follows \cite{viable-fr-lcdm,viable-fr-lcdm1}
\begin{equation}
f(R)=R-2\Lambda-\varpi (\frac{\Lambda}{R-4\Lambda})^{p_{+}-1}~ _{2}F_{1}[q_{+},p_{+}-1;r_{+};-\frac{\Lambda}{R-4\Lambda}],\label{fR-function}
\end{equation}
where $\varpi=\frac{3H_{0}^{2}\Omega_{\Lambda}D}{p_{+}-1}(\frac{\Omega_{m}}{\Omega_{\Lambda}})^{p_{+}}$, $_{2}F_{1}[a,b;c;z]$ is the Gaussian hypergeometric function with $q_{+}=\frac{1+\sqrt{73}}{12}$, $r_{+}=1+\frac{\sqrt{73}}{6}$ and $p_{+}=\frac{5+\sqrt{73}}{12}$. $D$ is the model parameter in this $f(R)$ modified gravity, which can relate to the current value $f_{R0}$ and the current value of the Compton wavelength $B_{0}$ by
\begin{equation}
f_{R0}=1+D\times _{2}F_{1}[q_{+},p_{+};r_{+};-\frac{\Omega_{\Lambda}}{\Omega_{m}}]
\end{equation}
\begin{equation}
B_{0}=\frac{2Dp_{+}}{\Omega_{m}^{2}\{1+D_{2}F_{1}[q_{+},p_{+};r_{+};-\frac{\Omega_{\Lambda}}{\Omega_{m}}]\}} \times \{\Omega_{\Lambda}\frac{q_{+}}{r_{+}}~ _{2}F_{1}[q_{+}+1,p_{+}+1;r_{+}+1;-\frac{\Omega_{\Lambda}}{\Omega_{m}}]
-\Omega_{m2}F_{1}[q_{+},p_{+};r_{+};-\frac{\Omega_{\Lambda}}{\Omega_{m}}]\},
\end{equation}
where the Compton wavelength is derived by $B=\frac{f_{RR}}{f_{R}}\frac{dR}{d\ln a}\frac{H}{dH/d\ln a}=
\frac{\partial f_{R}/\partial \ln a}{f_{R}}\frac{H}{\partial H/\partial \ln a}$.

Obviously,  Eq. (\ref{fR-function})  can partly realize the background expansion as that of the $\Lambda$CDM universe, while the cosmological perturbation behaviors in this $f(R)$ model are different from that in $\Lambda$CDM model.  Given that it is not natural by using two $f(R)$ functions to mimic one total $\Lambda$CDM universe,  in this paper  we consider our universe including two stages:  the early universe $a<0.02$ (including the radiation-dominate epoch and the early stage of the matter-dominate era) is described by the $\Lambda$CDM, and  the universe $a\geq 0.02$ (including the deep matter-dominate epoch and the late-time acceleration)  is depicted by the above viable $f(R)$ model.

\section{$\text{Cosmological perturbations in viable f(R) gravity theory producing  $\Lambda$CDM-background expansion}$}

The line element with the perturbation reads
\begin{equation}
ds^{2}=a^{2}[-(1+2\psi Y^{(s)})d\tau ^{2}+2BY_{i}^{(s)}d\tau dx^{i}+(1+2\phi Y^{(s)})\gamma_{ij}dx^{i}dx^{j}+\varepsilon Y_{ij}^{(s)}dx^{i}dx^{j}],
\end{equation}
where $\gamma_{ij}$ is the three-dimensional spatial metric in the spherical coordinate
\begin{eqnarray}
&&[\gamma_{ij}]=\left(\begin{array}{c}
\frac{1}{1-Kr^{2}}~~~~0~~~~~~~~0\\
~~0~~~~~~~r^{2}~~~~~~~~0\\
~~~~0~~~~~~~~0~~~~~~~~r^{2}\sin^{2}\theta
\end{array}\right),\label{gamma-ij}
\end{eqnarray}
 $(\triangle +k^{2})Y^{(s)}=0$,  $Y_{j}^{(s)}\equiv -\frac{1}{k}Y_{|j}^{(s)}$ and  $Y_{ij}^{(s)}\equiv \frac{1}{k^{2}}Y_{|ij}^{((s)}+\frac{1}{3}\gamma_{ij}Y^{((s)}$ are the scalar harmonic functions. Considering the synchronous gauge, we have $\psi= 0$, $B = 0$, $h_{L}=6\phi$ and $\eta_{T}=-(\phi+\varepsilon/6)$,
where $\eta_{T}=\frac{\delta R^{(3)}}{-4k^{2}+12K}=\frac{6\delta K}{-4k^{2}+12K}$ denotes the conformal 3-space curvature perturbation. The perturbed modified Einstein equations in $f(R)$ theory can be derived as follows \cite{xu-fr2-code}
\begin{equation}
(f_{R}\mathcal{H}+\frac{1}{2}f_{R}^{'})k\mathcal{Z}=\frac{\kappa^{2}}{2}a^{2}\delta\rho+f_{R}k^{2}\eta_{T}\beta_{2}-\frac{3}{2}\mathcal{H}\delta f_{R}^{'}-\frac{1}{2}\delta f_{R}k^{2}+\frac{3}{2}\mathcal{H}^{'}\delta f_{R}
\end{equation}
\begin{equation}
\frac{k^{2}}{3}f_{R}(\beta_{2}\sigma-\mathcal{Z})=\frac{\kappa}{2}a^{2}q+\frac{1}{2}k\delta f_{R}^{'}-\frac{1}{2}k\mathcal{H}\delta f_{R}
\end{equation}
\begin{equation}
\sigma^{'}+2\mathcal{H}\sigma+\frac{f_{R}^{'}}{f_{R}}\sigma=k\eta_{T}-\kappa^{2}a^{2}\frac{p\Pi}{f_{R}k}-k\frac{\delta f_{R}}{f_{R}}
\end{equation}
\begin{equation}
\mathcal{Z}^{'}+(\frac{1}{2}\frac{f_{R}^{'}}{f_{R}}+\mathcal{H})\mathcal{Z}=(-k\beta_{2}+\frac{k}{2}+\frac{3\mathcal{H}^{2}}{k})\frac{\delta f_{R}}{f_{R}}-\frac{\kappa^{2}a^{2}}{2kf_{R}}(\delta \rho+3\delta p)-\frac{3}{2}\frac{\delta f_{R}^{''}}{kf_{R}}
\end{equation}
where  $q=(\rho+p)v$, $\beta_{2}=\frac{k^{2}-3K}{k^{2}}$, $f_{R}=1+D a^{3p_{+}}\times_{2}F_{1}[q_{+},p_{+};r_{+};-a^{3}\frac{\Omega_{\Lambda}}{\Omega_{m}}]$, $\mathcal{H} = a^{'}/a$ is the conformal Hubble parameter, and superscript $'$ denotes the derivative with respect to conformal time. In addition, in the CAMB code, the curvature perturbations are characterized by $\mathcal{Z}=\frac{h_{L}^{'}}{2k}$ and $\sigma=\frac{(h_{L}+6\eta_{T})^{'}}{2k}$ with $\eta_{T}^{'}=\frac{k}{3}(\sigma-\mathcal{Z})$. The evolutional equation of the perturbed field $\delta f_{R}$ reads
\begin{equation}
\delta f_{R}^{''}+2\mathcal{H}\delta f_{R}^{'}+a^{2}(\frac{k^{2}}{a^{2}}+m^{2}_{f_{R}})\delta f_{R}=\frac{\kappa^{2}a^{2}}{3}(\delta\rho-3\delta p)-kf_{R}^{'}\mathcal{Z}.
\end{equation}

The source term of the CMB temperature anisotropy is described by
\begin{eqnarray}
S_{T}(\tau,k)&=&e^{-\varepsilon}(\alpha^{''}+\eta_{T}^{'})+g(\triangle_{T0}+2\alpha^{'}+\frac{v_{b}^{'}}{k}+\frac{\zeta}{12\sqrt{\beta_{2}}}
+\frac{\zeta^{''}}{4k^{2}\sqrt{\beta_{2}}})+g^{'}(\alpha+\frac{v_{b}}{k}+\frac{\zeta^{'}}{2k^{2}\sqrt{\beta_{2}}})
+\frac{1}{4}\frac{g^{''}\zeta}{k^{2}\sqrt{\beta_{2}}}\\\nonumber
&=&e^{-\varepsilon}(\frac{\sigma^{''}}{k}+\frac{k\sigma}{3}-\frac{k\mathcal{Z}}{3})+g(\triangle_{T0}+2\frac{\sigma^{'}}{k}+\frac{v_{b}^{'}}{k}
+\frac{\zeta}{12\sqrt{\beta_{2}}}+\frac{\zeta^{''}}{4k^{2}\sqrt{\beta_{2}}})+g^{'}(\frac{\sigma}{k}+\frac{v_{b}}{k}+\frac{\zeta^{'}}{2k^{2}\sqrt{\beta_{2}}})
+\frac{1}{4}\frac{g^{''}\zeta}{k^{2}\sqrt{\beta_{2}}}
\end{eqnarray}
where $g=-\dot{\varepsilon}e^{-\varepsilon}=an_{e}\sigma_{T}e^{-\varepsilon}$ is the visibility function and $\varepsilon$ is the optical depth. $\zeta$ is given by
$\zeta=(\frac{3}{4}I_{2}+\frac{9}{2}E_{2})$, where $I_{2}$,$E_{2}$ indicate the quadrupole of the photon intensity and the E-like polarization, respectively.

\section{$\text{Date fitting and results}$}

\subsection{$\text{Used data}$}

 In this section, we apply the cosmic data to constrain the above viable $f(R)$ model. The used data  are as follows.

 (1) The CMB temperature and polarization  information released by Planck 2015 \cite{Planck2015-1}: the high$-l$ $C_{l}^{TT}$ likelihood  (PlikTT), the high$-l$ $C_{l}^{EE}$  likelihood  (PlikEE), the high$-l$ $C_{l}^{TE}$  likelihood (PlikTE), the low$-l$ data and the lensing data.

 (2) The 10 datapoints of redshift space distortion (RSD): the RSD measurements from 6dFGS ($f\sigma_{8}(z = 0.067)=0.42\pm 0.06$) \cite{RSD1}, 2dFGRS ($f\sigma_{8}(z = 0.17)=0.51\pm 0.06$) \cite{RSD2}, WiggleZ ($f\sigma_{8}(z = 0.22)=0.42 \pm 0.07, f\sigma_{8}(z = 0.41)=0.45 \pm 0.04, f\sigma_{8}(z = 0.60)=0.43 \pm 0.04, f\sigma_{8}(z = 0.78)=0.38 \pm 0.04$) \cite{RSD3}, SDSS LRG DR7 ($f\sigma_{8}(z = 0.25)=0.39 \pm 0.05$, $f\sigma_{8}(z = 0.37)=0.43 \pm 0.04$) \cite{RSD4}, BOSS CMASS DR11 ($f\sigma_{8}(z = 0.57)=0.43 \pm 0.03$) \cite{RSD5}, and VIPERS ($f\sigma_{8}(z = 0.80)=0.47 \pm 0.08$) \cite{RSD6}. Here $f=\frac{d\ln D}{d\ln a}$, $D$ is the linear growth rate of matter fluctuations,  $\sigma_{8}$ is the RMS matter fluctuations in linear theory.  RSD reflects the coherent motions of galaxies, so it provides information about the formation of large-scale structure \cite{RSD7,xu-RSD1,xu-RSD2}.

 (3) The BAO data: the 6dFGS  \cite{bao1-6dFGS}, the SDSS-MGS \cite{bao2-SDSS-MGS}, the BOSSLOWZ BAO measurements of $D_{V}=r_{drag}$  \cite{bao2-SDSS-MGS} and the CMASS-DR11 anisotropic BAO  measurements  \cite{bao2-SDSS-MGS}. Since the WiggleZ volume partially overlaps that of the BOSSCMASS sample, we do not use the WiggleZ results in this paper.   6dFGS denotes the six-degree-Field Galaxy survey (6dFGS) at $z_{eff}=0.106$ \cite{bao1-6dFGS}, SDSS-MGS denotes the SDSS Main Galaxy Sample (MGS) at  $z_{eff}=0.15$ \cite{bao2-SDSS-MGS}, BOSSLOWZ denotes the Baryon Oscillation Spectroscopic Survey (BOSS) "LOWZ" at $z_{eff}=0.32$ \cite{bao2-SDSS-MGS}, and CMASS-DR11 denotes the BOSS CMASS at $z_{eff}=0.57$  \cite{bao2-SDSS-MGS}. The recent analysis of latter two BAO data use peculiar velocity field reconstructions to sharpen the BAO feature and reduce the errors on $D_{V}=r_{drag}$.  The point labelled BOSS CMASS at $z_{eff}=0.57$ shows $D_{V}=r_{drag}$ from the analysis of \cite{bao3}, updating the BOSS-DR9 analysis.

 (4) The supernova Ia (SN Ia) data from SDSS-II/SNLS3 joint light-curve analysis (JLA) \cite{sn,sn1}.

 The prior value of Hubble constant $H_{0}=100 h$ km s$^{-1}$ Mpc$^{-1}$ is usually taken in cosmic analysis, though there are hundreds of measurement value of $H_{0}$ and lots of them are mutually inconsistent\footnote{Measurements provide some different results on $H_{0}$, which are almost in the region $(60-80)$ $km s^{-1}Mpc^{-1}$, such as the higher values: $H_{0} = 74.3\pm 2.6$ $km s^{-1} Mpc^{-1}$ \cite{H0-high1},  the lower values: $H_{0} = 63.7\pm 2.3$ $km s^{-1} Mpc^{-1}$ \cite{H0-low1} and the “concordance” value: $H_{0} = 69.6\pm 0.7$ $km s^{-1} Mpc^{-1}$ \cite{H0-middle1}, etc. For other measurement values of $H_{0}$, one can see Ref.\cite{H01,H02,H03,H04,H05,H06,H07}}.   Ref. \cite{H0-prior-chen} points out that the prior value of the $H_{0}$  affects cosmological parameter estimation, but not very significantly.  Here we take the HST prior, $H_{0} = 73.8 \pm 2.4$ $km s^{-1}Mpc^{-1}$ \cite{HST}.

\subsection{$\text{Constraint on  neutrino mass and  the base parameters in viable $f(R)$ model producing  $\Lambda$CDM expansion}$}

\begin{figure}[!htbp]
  \includegraphics[width=7.5cm]{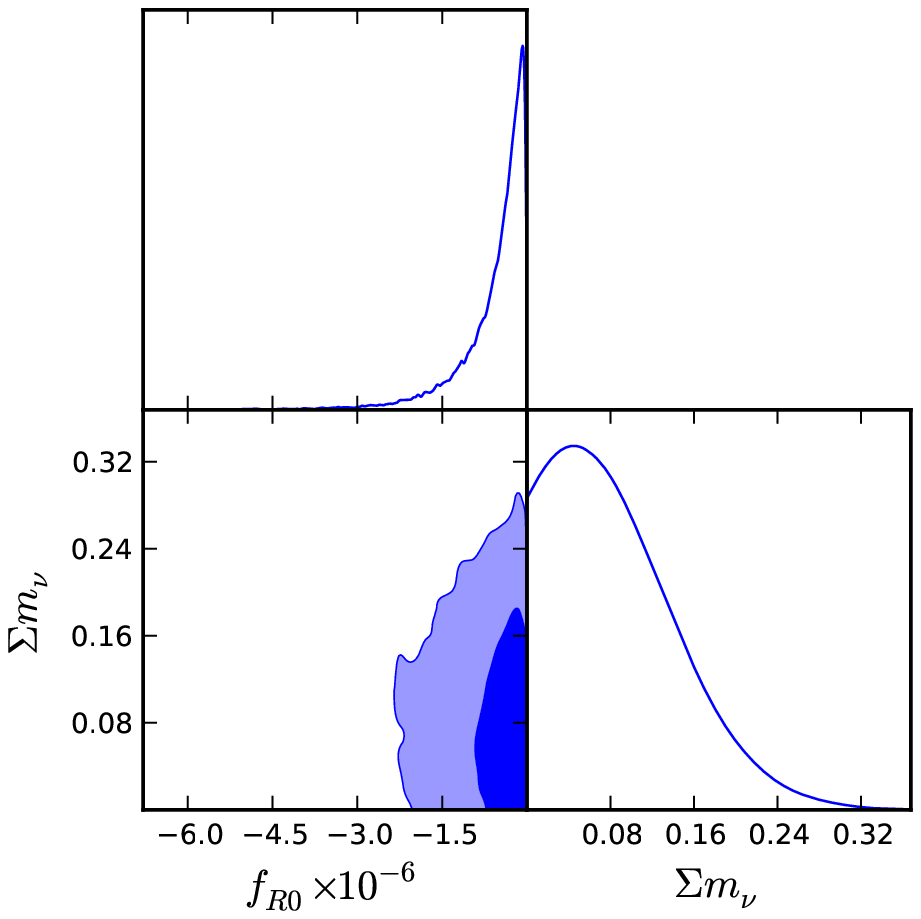}
  \includegraphics[width=7.5cm]{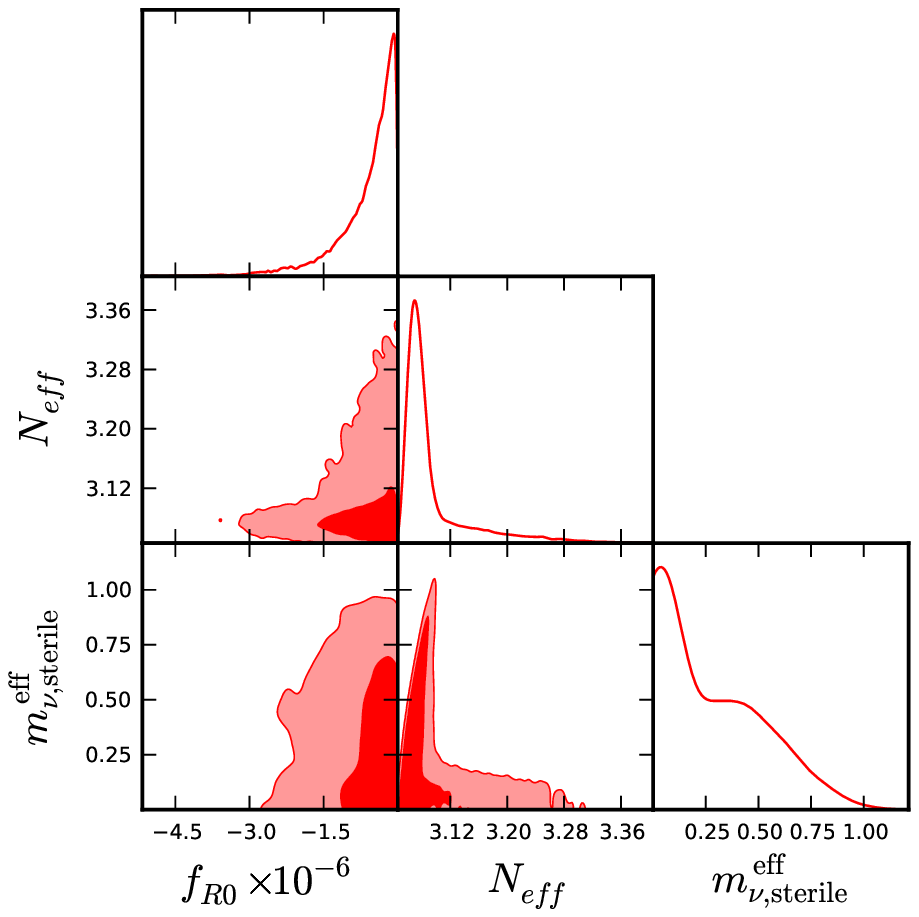}
  \caption{The contours of model parameters in viable $f(R)$ theory with massive neutrino  by fitting the Planck TT, TE, EE+lowP and the low-redshift data: Planck lesning+RSD+BAO+JLA.}\label{figure-fr-neff-mnu}
\end{figure}

\begingroup
\squeezetable
\begin{center}
\begin{table}
\begin{tabular}{c|c|c}
\hline\hline
Parameters  &  active  & sterile  \\\hline
{\boldmath$\Sigma m_\nu   $} & $< 0.202                   $  &  ---- \\
{\boldmath$m_{\nu,{\rm{sterile}}}^{\rm{eff}}$} & ----        &  $< 0.757     $\\
{\boldmath$N_{eff}        $} &   ----                        &  $< 3.22       $  \\
{\boldmath$f_{R0}\times 10^{-6}         $} & $> -1.89                   $  &  $> -2.02      $  \\\hline
{\boldmath$\Omega_b h^2   $} & $0.02233^{+0.00028}_{-0.00028}$  & $0.02228^{+0.00031}_{-0.00029}$ \\
{\boldmath$\Omega_c h^2   $} & $0.1178^{+0.0022}_{-0.0022}$     &  $0.1147^{+0.0063}_{-0.0068}$\\
{\boldmath$100\theta_{MC} $} & $1.04090^{+0.00060}_{-0.00059}$  &  $1.04096^{+0.00063}_{-0.00065}$\\
{\boldmath$\tau           $} & $0.053^{+0.026}_{-0.027}   $  &  $0.060^{+0.028}_{-0.028}$ \\
{\boldmath${\rm{ln}}(10^{10} A_s)$} & $3.035^{+0.048}_{-0.050}   $  & $3.049^{+0.052}_{-0.053}   $ \\
{\boldmath$n_s            $} & $0.9683^{+0.0080}_{-0.0079}$  &  $0.9713^{+0.0097}_{-0.0087}$ \\\hline
\end{tabular}
\caption{The 95\% confidence level of basic parameters in viable $f(R)$ model with the massive neutrino by fitting the Planck TT, TE, EE+lowP and the low-redshift data: Planck lesning+RSD+BAO+JLA.}\label{tab:fR-base}
\end{table}
\end{center}
\endgroup

Constraints on neutrino mass in $\Lambda$CDM model or in dynamical dark energy models or in $f(R)$ theory have been discussed in some references \cite{Planck2015-1,Zhang-1511-02651-prd-mnu,1308-5870-prl-sterile,Geng-1411-3813-plb-mnu,Zhang-1502-01136-plb-sterile,He-1307-4876-prd-mnu,1412-5239-jcap-fR-sterile}.
 Given that the constraints on  $\Sigma m_{\nu}$ (or $ m_{\nu, eff}^{sterile}$) are model-dependent,  we fit the cosmic data to limit the  mass of neutrinos in above viable $f(R)$ model by using the MCMC method \cite{cosmomc,cosmomc1,cosmomc2,cosmomc3,cosmomc4,cosmomc5}. Obviously,  extra parameters $f_{R0}$ and $\Sigma m_\nu$ (or $ m_{\nu, eff}^{sterile}$  with the required $N_{eff}$) are added,  relative to the base $\Lambda$CDM model. Table  \ref{tab:fR-base} lists the 95\% limits of basic parameters in $f(R)$ model. One can see the upper bound on the mass of active neutrino $\Sigma m_{\nu}< 0.202$, which is comparable with other results. For example, adding  a single free parameter $\Sigma m_{\nu}$ to the base $\Lambda$CDM model, fitting different data gives $\Sigma m_{\nu}<0.177$ eV  \cite{Zhang-1511-02651-prd-mnu}, $\Sigma m_{\nu}<0.17$ eV \cite{Planck2015-1} or $\Sigma m_{\nu}<0.254$ eV \cite{1308-5870-prl-sterile};  Adding $\Sigma m_{\nu}$ to the dynamical DE model, fitting cosmic data gives $\Sigma m_{\nu}<0.304$ eV in $w$CDM model \cite{Zhang-1511-02651-prd-mnu} or $\Sigma m_{\nu}<0.113$ eV in holographic DE model  \cite{Zhang-1511-02651-prd-mnu}; And  $\Sigma m_{\nu}<0.451$ eV and $\Sigma m_{\nu}<0.214$ eV are given in Starobinsky $f(R)$ model and exponential $f(R)$ model \cite{Geng-1411-3813-plb-mnu}, respectively.\footnote{Constraint on total mass of active neutrino are also investigated with including an additional free parameter $N_{eff}$ in theoretical model, for example,  $\Sigma m_\nu<0.826$  with  $N_{eff}=3.49_{-0.73}^{+0.71}$ \cite{He-1307-4876-prd-mnu} and $\Sigma m_\nu= 0.533_{- 0.411}^{+0.254}$  with  $N_{eff}=3.78_{-0.64}^{+0.84}$ \cite{Geng-1411-3813-plb-mnu} are given in the $f(R)$ models.} Table  \ref{tab:fR-base} also exhibits the constraint result $m_{\nu, eff}^{sterile}<0.757$ with  $N_{eff}<3.22$ for sterile neutrino case  in viable $f(R)$ model. One can compare this results with other ones.  For example, fitting the different cosmic data gives  $ m_{\nu, eff}^{sterile}<0.52$ eV with $N_{eff}<3.7$ \cite{Planck2015-1},  $m_{\nu, eff}^{sterile}<0.479$ eV with $\Delta N_{eff}=<0.98$  \cite{1308-5870-prl-sterile}, or $ m_{\nu, eff}^{sterile}<0.43$ with $N_{eff}<3.96$ in $\Lambda$CDM model  \cite{Zhang-1502-01136-plb-sterile}, and  $ m_{\nu, eff}^{sterile}<0.61$ with $N_{eff}<3.95$ in $f(R)$ model \cite{Zhang-1502-01136-plb-sterile}. Obviously, a higher upper limit on $ m_{\nu, eff}^{sterile}$ and a lower limit on $N_{eff}$ are obtained in our study. Some inconsistent  results on  sterile neutrino mass can also be found, for example, the sterile neutrino mass $0.47 eV < m_{\nu,sterile}^{eff}<1 eV$ ($2\sigma$) is given in a $f(R)$ model  and  $0.45eV < m_{\nu,sterile}^{eff}< 0.92 eV$  is given in $\Lambda$CDM model \cite{1412-5239-jcap-fR-sterile}, or  the active neutrino mass $\sum m_{\nu}=0.35\pm 0.10$ is presented in $\Lambda$CDM model  \cite{massive-neutrinos1}. The constraint results  on model parameter in viable $f(R)$ theory are $f_{R0}\times 10^{-6}> -1.89$ for active neutrino case and $f_{R0}\times 10^{-6}> -2.02$ for sterile neutrino case at \%95 limit.  Though the fitting results on  $f_{R0}$ are affected by the additional parameters $\Sigma m_\nu$ (or $m_{\nu,sterile}^{eff}$ with $N_{eff}$), for using the Planck 2015 data in this paper it has the more stringent constraint than result given by Ref. \cite{xu-1411-4353-prd-fR0}:  $f_{R0}\times 10^{-6}=-2.58^{+2.14}_{-0.58}$ in $1\sigma$ regions.

Table \ref{tab:fR-base} also lists the values of  six basic cosmological parameters.  $\Omega_{b}h^2$ is the current baryon density,  $\Omega_{c}h^2$ is the cold dark matter density at present, $\theta_{MC}$ denotes the approximation to $r_{*}/D_{A}$, $\tau$ presents the Thomson scattering optical depth due to reionization, ${\rm{ln}}(10^{10} A_s)$ is the Log power of the primordial curvature perturbations, and $n_s$ is the scalar spectrum power-law index. From  table \ref{tab:fR-base} and figure \ref{figure-fr-base}, it can be seen that the neutrino properties much more affect the fitting value of  cold dark matter density than fitting values of other parameters. This results could be interpreted as follow. Since the massive neutrinos  are considered as one kind of dark matter in universe, the mass of neutrino (active or sterile) would directly affect the dimensionless energy density of dark matter. According to the constraint results on $\Omega_{c}h^{2}$ and $\Omega_{\nu}h^{2}$, one can see that the larger uncertainty of $\Omega_{c}h^{2}$ value is caused by the looser constraint on the dimensionless energy density of sterile neutrino $\Omega_{\nu}h^{2}$, which maybe reflects the less information on sterile neutrino from cosmic observations. However, the constraint on $\Omega_{c}h^{2}$ is more strict for the active-neutrino case, since  the  constraint on the dimensionless energy density of active neutrino $\Omega_{\nu}h^{2}$ is tighter than the case of sterile neutrino, which maybe reflects the more information on the active neutrino  from cosmic observations. Except $\Omega_{c}h^{2}$, other basic parameters in table \ref{tab:fR-base} are not directly related to neutrino density, so affecting on fitting values of other basic parameters from the neutrino characters are smaller.

\subsection{$\text{Constraint on derived  parameters in viable $f(R)$ model  producing  $\Lambda$CDM expansion}$}

\begin{figure}[!htbp]
  \includegraphics[width=12cm]{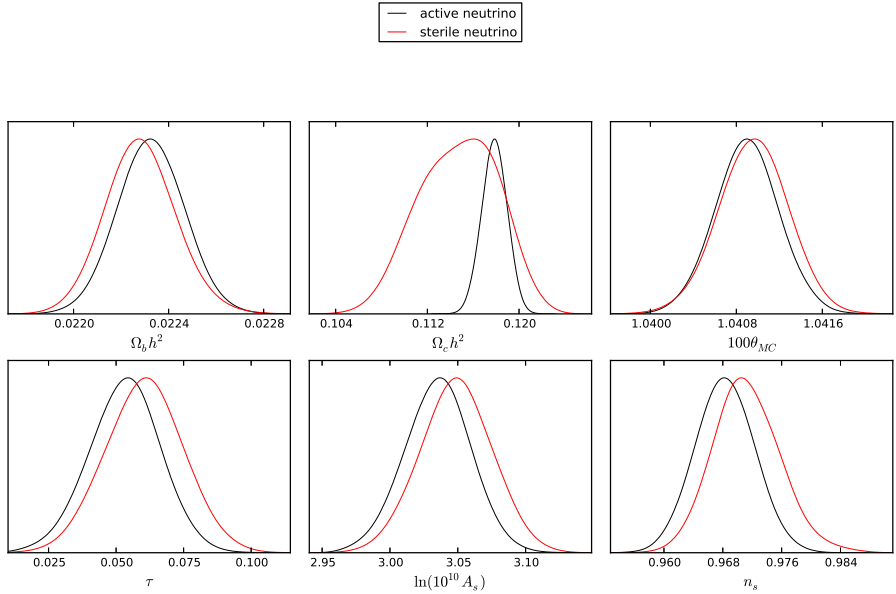}
  \caption{The $1-D$ distributions of basic cosmological parameters in viable $f(R)$ model with massive neutrino.}\label{figure-fr-base}
\end{figure}
%%%%%%%%%%%%%%%%%%%%%%%%%%%%%%%%%%%%%%%%%%%%%%%%%%%%%%%%
\begingroup
\squeezetable
\begin{center}
\begin{table}
\begin{tabular}{c|c|c|c|c|c}
\hline\hline
Parameters                      &  active                   & sterile        & Parameters                      & active                   & sterile  \\\hline
$H_0                       $ & $67.9^{+1.1}_{-1.1}        $ & $68.4^{+1.1}_{-0.99}       $ &
%$\Omega_\Lambda            $ & $0.694^{+0.013}_{-0.014}   $ & $0.699^{+0.013}_{-0.012}   $\\
$\sigma_8                  $ & $0.813^{+0.023}_{-0.023}   $ & $0.811^{+0.023}_{-0.022}   $\\\hline
$\Omega_m                  $ & $0.306^{+0.014}_{-0.013}   $ & $0.301^{+0.012}_{-0.013}   $&
$\Omega_m h^2              $ & $0.1411^{+0.0019}_{-0.0019}$ & $0.1410^{+0.0032}_{-0.0027}$ \\
$\Omega_m h^3              $ & $0.09579^{+0.00085}_{-0.00096}$ & $0.0964^{+0.0026}_{-0.0013}$&
$\sigma_8 \Omega_m^{0.5}   $ & $0.450^{+0.012}_{-0.012}   $ & $0.445^{+0.014}_{-0.014}   $ \\
$\sigma_8 \Omega_m^{0.25}  $ & $0.605^{+0.015}_{-0.015}   $ & $0.601^{+0.017}_{-0.017}   $&
$z_{\rm re}                $ & $7.5^{+2.6}_{-2.8}         $ & $8.2^{+2.6}_{-3.0}         $ \\
$10^9 A_s                  $ & $2.08^{+0.10}_{-0.10}      $ & $2.11^{+0.11}_{-0.11}      $&
$10^9 A_s e^{-2\tau}       $ & $1.870^{+0.023}_{-0.022}   $ & $1.869^{+0.025}_{-0.023}   $ \\
$t_{0} ({\rm{Gyr}})        $ & $13.803^{+0.068}_{-0.061}  $ & $13.762^{+0.074}_{-0.12}   $&
$z_*                       $ & $1089.79^{+0.47}_{-0.46}   $ & $1089.89^{+0.49}_{-0.48}   $ \\
$r_*                       $ & $145.02^{+0.51}_{-0.49}    $ & $144.87^{+0.88}_{-1.4}     $ &
$100\theta_*               $ & $1.04111^{+0.00060}_{-0.00059}$ & $1.04113^{+0.00064}_{-0.00069}$\\
$z_{\rm{drag}}             $ & $1059.68^{+0.60}_{-0.58}   $ & $1059.61^{+0.78}_{-0.72}   $&
$r_{\rm{drag}}             $ & $147.71^{+0.52}_{-0.50}    $ & $147.57^{+0.91}_{-1.5}     $ \\
$k_{\rm D}                 $ & $0.14018^{+0.00058}_{-0.00060}$ & $0.1401^{+0.0012}_{-0.00092}$&
$z_{\rm{eq}}               $ & $3350^{+49}_{-50}          $ & $3252^{+120}_{-150}        $ \\
$k_{\rm{eq}}               $ & $0.01022^{+0.00015}_{-0.00015}$ & $0.01001^{+0.00035}_{-0.00038}$ &
$100\theta_{\rm{s,eq}}     $ & $0.4543^{+0.0050}_{-0.0048}$ & $0.465^{+0.017}_{-0.013}   $ \\\hline
$\log_{10}(B_{0})          $ & $-5.74^{+0.92}_{-1.00}      $ & $-5.69^{+0.90}_{-1.00}      $&
$\Omega_\nu h^2            $ & $0.00094^{+0.0013}_{-0.00097}$ & $0.0039^{+0.0048}_{-0.0034}$ \\
$Y_P                       $ & $0.24537^{+0.00012}_{-0.00013}$ & $0.2460^{+0.0018}_{-0.00071}$ &&& \\
 \hline\hline
\end{tabular}
\caption{The 95\% confidence level of derived parameters in viable $f(R)$ model with the massive neutrino by fitting the Planck TT, TE, EE+lowP and the low-redshift data: Planck lesning+RSD+BAO+JLA.}\label{tab:derived}
\end{table}
\end{center}
\endgroup

The values of interested derived parameters are calculated and listed in table \ref{tab:derived}. It includes  20 parameters listed in table 4 of Ref. \cite{Planck2015-1} and 3 other parameters $[\log_{10}(B_{0}),\Omega_{\nu}h^{2}, Y_{p}]$. Concretely, $\Omega_{m}$ is the current dimensionless matter density, $z_{re}$ is the redshift at which universe is half reionized,  $t_{0}$ denotes  the age of the universe today (in Gyr), $z_{*}$ denotes the redshift for which the optical depth equals unity, $r_{*}$ denotes the comoving size of the sound horizon at $z=z_{*}$, $\theta_{*}$ denotes the  angular size of sound horizon at $z=z_{*} (r_{*}/D_{A})$, $z_{drag}$ denotes the redshift at which baryon-drag optical depth equals unity, $r_{drag}$ denotes the comoving size of the sound horizon at $z=z_{drag}$, $k_{D}$ denotes the characteristic damping comoving wavenumber (Mpc$^{-1}$), $z_{eq}$ denotes the redshift of matter-radiation equality, $\Omega_{\nu}h^2$ is the neutrino density, $Y_{p}$ denotes the fraction of baryonic mass in helium. Obviously, from table  \ref{tab:derived} we can see that the constraint results on  $H_{0}$  and $\sigma_{8}$  are: $H_{0}=67.9^{+1.1}_{-1.1}$ and $\sigma_{8}=0.813^{+0.023}_{-0.023}$ for active neutrino case, $H_{0}=68.4^{+1.1}_{-0.99} $ and $\sigma_{8}=0.811^{+0.023}_{-0.022}$ for sterile neutrino case, which are compatible with results given by \cite{Zhang-1511-02651-prd-mnu}. For these constraint results on $H_{0}$, it is also shown that the tension between direct and CMB measurements of $H_0$ gets slightly weaker in our considered model than that in the base $\Lambda$CDM model, where $H_{0}=67.6\pm 0.6$ is given by Ref. \cite{Planck2015-1}. In addition, it is found from Fig. \ref{figure-fr-derived-1D} that the neutrino properties much affect the fitting value of parameters: $z_{eq}$, $k_{eq}$, $100\theta_{s,eq}$, $\Omega_{\nu}h^{2}$ and $Y_{p}$, which  could be partly explained by the dependency of the parameters on the cold dark matter density and might be useful for  testing the neutrino properties in experiments. The values of $\sigma_{8}$ in viable $f(R)$ model are almost the same for the cases of different-species neutrino, and the same result is also suitable for the  parameters: $f\sigma_{8}$, $A_s e^{-2\tau}$ and $\theta_{*}$, as exhibited in Fig. \ref{figure-fr-derived-1D} and Fig. \ref{figure-fsigma8}.

\begin{figure}[!htbp]
  \includegraphics[width=18cm]{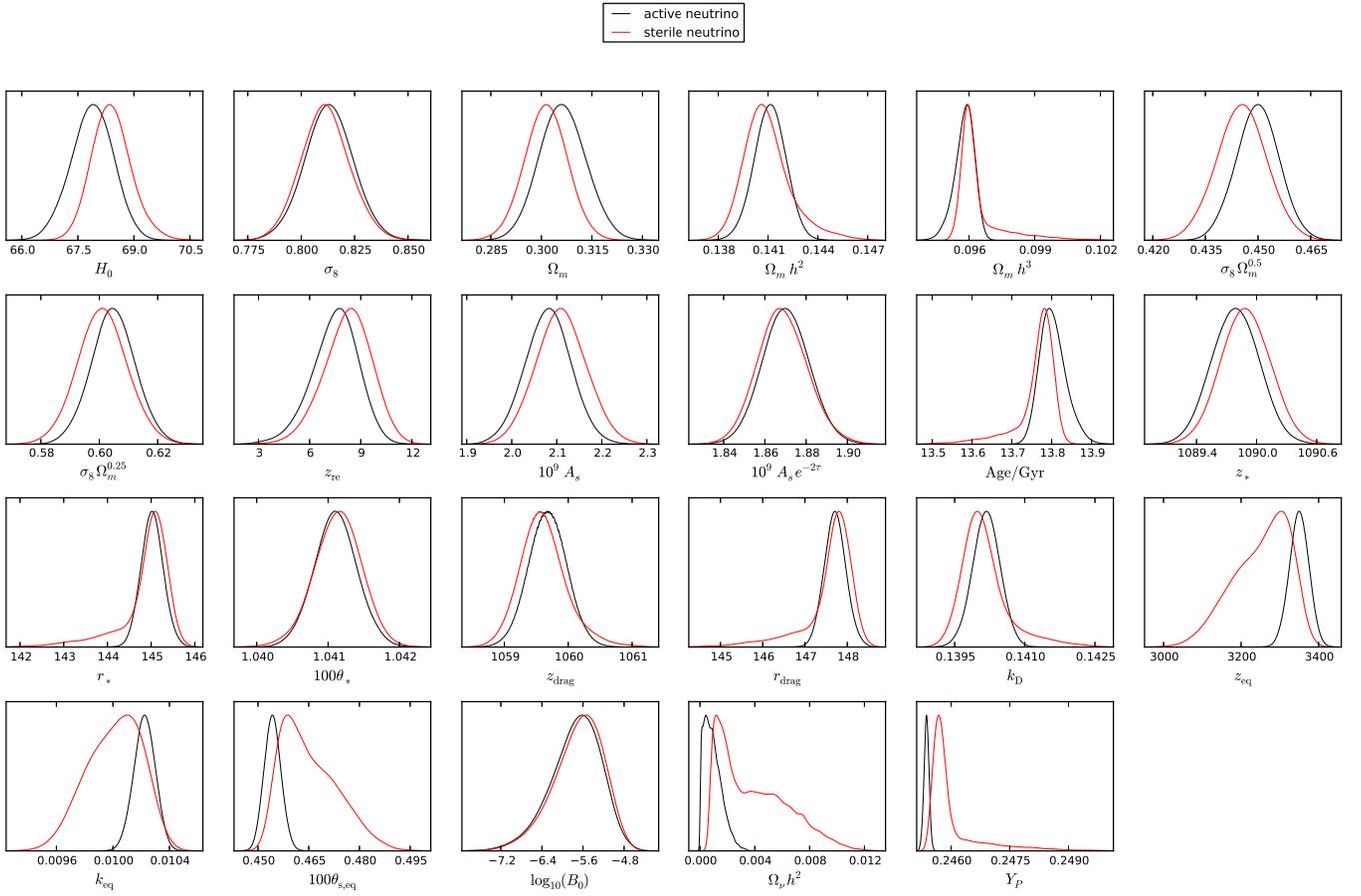}
  \caption{The $1-D$ distributions of derived cosmological parameters in viable $f(R)$ model with massive  neutrino.}\label{figure-fr-derived-1D}
\end{figure}

\begin{figure}[!htbp]
  \includegraphics[width=6.5cm]{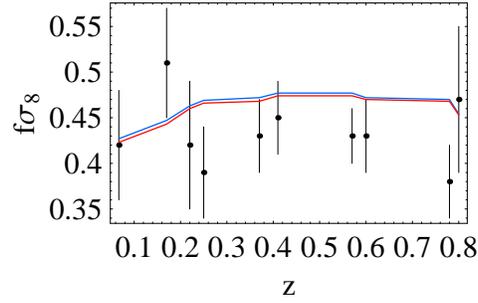}
  \caption{The values of $f\sigma_{8}$ calculated in the viable $f(R)$ model with the massive neutrino, where blue line denotes the active-neutrino case, and red line denotes the sterile-neutrino case. Ten dots with error bar denote the observational datapoints from Refs. \cite{RSD1,RSD2,RSD3,RSD4,RSD5,RSD6}}.\label{figure-fsigma8}
\end{figure}
\

\section{$\text{Conclusion }$}
  Tensions between several observations  were found  recently. The studies on tensions are important, since they are useful to  search new physics.  The massive neutrinos are introduced in cosmological models to solve the tensions concerning the inconsistent values of $H_{0}$ (or $\sigma_{8}$).  Investigating other scenarios to solve these tensions and restricting the mass of neutrinos in different scenarios are significative. Given that several forms of  $f(R)$ are found to be nonphysical, we study  the viable $f(R)$ gravity with the massive neutrinos  in this paper. We fit the current observational data: Planck-2015 CMB, RSD, BAO  and SNIa to constrain the mass of  neutrinos in viable $f(R)$ theory. The constraint results at 95\% confidence level are: $\Sigma m_\nu<0.202$ eV for the active neutrino case and $m_{\nu, sterile}^{eff}<0.757$ eV with $N_{eff}<3.22$ for the sterile neutrino case, which are comparable with some other results. For the effects by the mass of neutrinos, the constraint results  on model parameter become $f_{R0}\times 10^{-6}> -1.89$  and $f_{R0}\times 10^{-6}> -2.02$ for two cases, respectively. It is also shown that the fitting values of several parameters much depend on the  neutrino properties, such as the cold dark matter density $\Omega_{c}h^{2}$, the cosmological quantities at matter-radiation equality: $z_{eq}$, $k_{eq}$ and $100\theta_{s,eq}$, the neutrino density  $\Omega_{\nu}h^{2}$ and the fraction of baryonic mass in helium $Y_{p}$. At last, the constraint result shows that  the tension between direct and CMB measurements of $H_0$ gets slightly weaker in the viable $f(R)$ model than that in the base $\Lambda$CDM model.

 \textbf{\ Acknowledgments }
 We thank the Professor Lixin Xu for his useful discussion on this paper. The research work is supported by   the National Natural Science Foundation of China (11645003,11475143,11575075).

\end{document}